\documentclass[reprint,amssymb,superscriptaddress]{revtex4-2} 
\usepackage{amsmath} 
\usepackage{amssymb} 
\usepackage{amsfonts} 
\usepackage{bm} 
\usepackage{amssymb} 
\usepackage{graphicx} 
\usepackage{dcolumn} 
\usepackage{txfonts} 
\usepackage{makeidx}
\usepackage{color}
\usepackage{mathtools}
\usepackage{threeparttable}
\usepackage[linkcolor=blue,anchorcolor=black,citecolor=blue,colorlinks=true]{hyperref}
\usepackage{breakurl}
\usepackage{multirow}
\usepackage{array}
\usepackage{accents}
\usepackage{makecell}

\begin{document}
\title{Experimental verification of the optimal fingerprint method in a Spin Resonance System: Implications for Complex Systems}
\author{Jinbo Hu}
\thanks{These authors contributed equally to this work}
\affiliation{Beijing Computational Science Research Center, Beijing 100193, People's Republic of China}
\author{Hong Yuan}
\thanks{These authors contributed equally to this work}
\affiliation{Graduate School of CAEP, Beijing 100193, People's Republic of China}
\author{Letian Chen}
\affiliation{Graduate School of CAEP, Beijing 100193, People's Republic of China}
\affiliation{Department of Mathematics and Centre of Complexity Science, Imperial College London, London SW7 2BZ, United Kingdom}
\author{Nan Zhao}
\email{nzhao@csrc.ac.cn}
\affiliation{Beijing Computational Science Research Center, Beijing 100193, People's Republic of China}
\author{C.P. Sun}
\email{suncp@gscaep.ac.cn}
\affiliation{Graduate School of CAEP, Beijing 100193, People's Republic of China}

\begin{abstract}
The optimal fingerprint method (OFM) serves as a potent approach for detecting and attributing climate change. 
However, direct experimental validation remains challenging due to the system's inherent complexity.
Here, we experimentally validate this method using a precisely controlled magnetic resonance system of spins, which serves as a minimal physical analog of forced noise response. 
Based on linear response theory (LRT), we derived the system's Green's function from spin projection noise measurements and successfully applied it to attribute the magnetic fields, yielding excellent agreement with predictions.
Furthermore, our measurements confirm the existence of an optimal detection direction that maximizes the signal-to-noise ratio, a key theoretical prediction underlying the OFM. 
This work serves as a laboratory demonstration of LRT and OFM in detection and attribution (DA) studies, aiming to connect theoretical models with experimental observations. 
These findings may provide useful references for climate change science and its broader interdisciplinary applications. in ecosystems, finance, social sciences, quantum sensing, and beyond.
\end{abstract}

\maketitle
\section{Introduction} 
The Earth's climate is indispensable for human survival, drawing significant attention to research due to its direct impact on both present living conditions and future development.  
Understanding its evolution is vital for formulating effective policies and strategies to cope with its effects. 
However, studying climate change comes with its own set of challenges.

The system's complexity, characterized by an array of interacting components, complicates straightforward mathematical representation. 
For instance, the chaotic nature of temperature evolution, influenced by numerous factors with complex mechanisms, challenges direct modelling. 
Nevertheless, the scientific community has made substantial progress in this arena \cite{2_Climate_Models0}. 
Researchers have devoted considerable effort towards the development of sophisticated models, including general circulation models (GCMs) \cite{3_GCMs1,4_GCMs2,5_GCMs_Manabe,6_GCMs4} and stochastic climate models (SCMs) \cite{7_SCMs1,8_SCMs2,9_SCMs3}. 
These models help us understand how the system responds to both natural events and human activities. 

However, given that climatology is an observational science, these models are not amenable to direct experimental verification; instead, they can only be validated against observational data.
This raises another pressing issue: how can human-induced changes be reliably distinguished from natural variability \cite{10_Methods2}. 
The climate observations are typically modeled numerically by the error-in-variables (EIV) model \cite{hannart_optimal_nodate}.
The observed data $\boldsymbol{\Psi}$ (an $S$-dimensional vector consisting of a series of $Y_k$) is usually assumed to be a linear superposition of 
the expected signal $\boldsymbol{\Psi}^{\rm s}$, namely, the forced deterministic one by a series of forcing $\tilde{X}^p_k$ ($p = 1,\cdots,M$) , and the noise $\delta\boldsymbol{\Psi}$ (a vector consisting of
a series of $\mathcal{R}_k$) of the natural system
\begin{equation}
    \boldsymbol{\Psi} = \boldsymbol{\Psi}^{\rm s} + \delta\boldsymbol{\Psi}.
\end{equation}
The linear combination relationship among  $Y_k$, $\tilde{X}^p$, and $\mathcal{R}_k$ can be expressed as
\begin{equation}
    Y_k = \sum_{p = 1}^{M} \tilde{X}_k^p\beta_p + \mathcal{R}_k, ~ k = 1,\cdots,S,
\end{equation}
Then seeking the optimal solution in the parameters $\beta_p$, that is,  how to separate the environmental noise $\mathcal{R}_k$ 
from the numerous climate observation variables $Y_k$ and obtain the human activities $\tilde{X}_k^p$ that are of interest, has become a key issue of concern \cite{hannart_optimal_nodate}.

For this purpose, Hasselmann introduced the optimal fingerprinting method (OFM), mathematically represented by the product of the expected signal pattern and the inverse of the variability covariance matrix \cite{10_Methods2,11_Methods1,12_Methods3}. 
Geometrically, the fingerprint represents a direction minimizing the influence of high-noise components, facilitating the detection of anthropogenic signals. 
Beyond detection, this method is also crucial to identifying the attribution to specific causes \cite{Benjamin2013,stott_human_2004,gillett_detection_2003,16_Application1,Barnett2008,19_Application4,20_importance1}. 
Based on this method, scientists have concluded that the human activities indeed impact the climate changes with high confidence \cite{21_masson2021ipcc}.
Recent studies have explored the application of response theory to gain dynamical and physical insights into the detection and attribution of climate signals \cite{lucarini_detecting_2024}. 
They find that the Green's function from linear response theory (LRT) provides accurate attribution in linear regimes. 
This finding establishes a robust theoretical foundation for OFM applied in fields such as ecosystems, quantitative social science, and finance.

Despite substantial theoretical and numerical evidence \cite{13_attribution0,14_attribution1,15_Application0,17_Application2,18_Application3} 
supporting the validity and superiority of the optimal fingerprint method, a direct, quantitative experimental validation is notably absent.
Unfortunately, the inherent complexity of systems once again poses significant challenges to the experimental validation of the optimal fingerprinting method. 
For instance, this method relies on certain assumptions such as the linear superposition of external forces and internal variability effects. 
The intricate nature of the systems undoubtedly renders the experimental verification of these conditions highly difficult.  
Despite these obstacles, the method's pivotal role in climate prediction and the profound implications of its conclusions underscore the necessity for its experimental validation.

In this study, we propose an experimentally verifiable system for demonstration. 
Specifically, we consider a {\it spin resonance system} with noise.
Atomic spins are subjected in a well-controlled magnetic field (the external force) and a stochastic magnetic field (the noise). 
With the adiabatic approximation and rotating wave approximation, the noise-driven part and the external-force-driven part of the spin signal are separated. 
Through LRT analysis, we determined the relationship between the external-force-driven part of the spin signal and the applied magnetic field drive.
The theoretical results are consistent with measurement data.
In addition, by combining the noise-driven part spin signal, we theoretically calculate the corresponding optimal direction, the fingerprint, under the restriction of maximizing the signal-to-noise ratio. 
The measured optimal direction matches the theoretical predictions, thereby confirming the effectiveness of the optimal fingerprint method in signal extraction and attribution issues,
and providing a solid experimental foundation for the further application of the optimal fingerprint method in DA studies.

This paper is organized as follows. In Sec.\ref{section_OFM} we give a brief introduction to the optimal fingerprint method. 
Section.\ref{section_LRT} presents how the LRT works in the DA studies.
Section.\ref{section_OFMINSPIN} presents the results of the optimal fingerprint method applied in spin systems. 
In Sec.\ref{section_experiment} we introduce our experimental setup and measurement results, which is consistent with our theoretical analysis.
Finally, the conclusion is presented in in Sec.\ref{section_conclusion}.
Also this study is supported by three appendices: 
Appendix~\ref{section_theory_ap} provided a theoretical analysis about the noise distribution in the spin system.
Appendix~\ref{section_GreenFunction_ap} gives a detailed derivation of the Green's function in spin systems and the analytical form of spin projection noise obtained using the fluctuation dissipation theorem.
Appendix~\ref{section_experiment_ap} gives the method about how we control the experiment condition and how we get our experiment data.

\section{Optimal fingerprint method}
\label{section_OFM}
The optimal fingerprint is given by the product of the expected signal $\boldsymbol{\Psi}^{\rm s}$ and the inverse covariance matrix of the noise $\delta\boldsymbol{\Psi}$.
By applying fingerprints to observed data (geometrically, it equal to project the signal along a certain measurement direction), a detection variable with the maximum signal-to-noise ratio can be obtained.
Given an arbitrary vector $\mathbf{n}$ (not necessarily normalized,, and referred to as the {\it measurement direction} hereafter), the square of the signal-to-noise ratio (SNR) along the direction of $\mathbf{n}$ is defined as
\begin{equation}
    \label{Eq:SNRtheorydefine}
R^2({\mathbf{n}}) \equiv \frac{(\boldsymbol{\Psi}^{\rm s}\cdot \mathbf{n})^2}{\left\langle\left(\delta\boldsymbol{\Psi}\cdot \mathbf{n}\right)^2\right\rangle}. 
\end{equation}
Let $\hat{n}_s$ be a unit-length vector denoting the direction of the signal, $\boldsymbol{\Psi}^{\rm s} = \Psi_{\rm s}\hat{n}^{\rm s}$
with $\Psi^{\rm s}=\vert \boldsymbol{\Psi}^{\rm s}\vert$ being the magnitude of the signal,
the signal $\boldsymbol{\Psi}^{\rm s}$ can be detected with high level of confidence 
as long as the SNR along $\hat{\mathbf{n}}^{\rm s}$ is sufficiently large, i.e., $R^2(\hat{\mathbf{n}}^{\rm s}) \gg 1$.

\begin{figure}
    \includegraphics[scale=0.6]{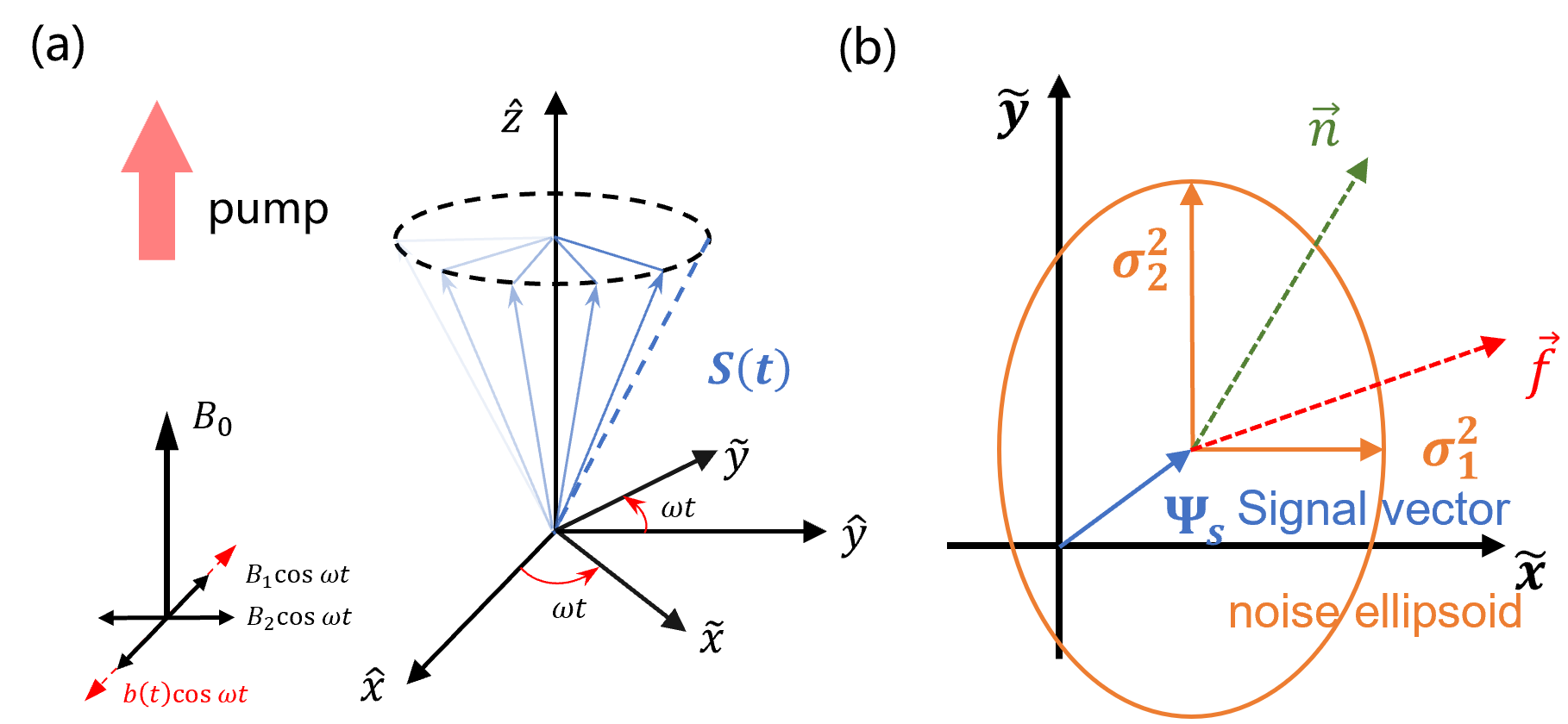}
    \caption{\label{fig:illustration_of_fingerprint}An illustration of the spin dynamics evolution and the fingerprint method.
        (a) The evolution of the spin driving by a resonant transverse magnetic field. 
        The spin vector precess (the blue arrows) in the main magnetic field $B_0$ with frequency $\omega_0 = \gamma B_0$.
        The spin components can be observed in either the laboratory frame (the $\hat{x}-\hat{y}$ frame) or the rotating frame (the $\tilde{x}-\tilde{y}$ frame).
        (b) The optimal fingerprint theory in the spin system.
        The spin signal $\boldsymbol{\Psi}$ is represented by a two-dimensional vector (the blue arrow) in the rotating frame.
        The orange ellipsoid shows the fluctuation of spin generated by the noise field.
        Both the signal and noise are projected to the detection direction $\mathbf{n}$ to estimate the SNR, which reaches the maximum in the optimal direction $\mathbf{f}$.
    }
\end{figure}

However, for a natural system, it is almost impossible to distinguish the expected signal from the noise, i.e., the expected signal is undetectable, without using some filtering techniques to eliminate unavailing information.
Hasselmann developed the optimal fingerprint method to detect the signal from noise.
The optimal fingerprint $\mathbf{f}$ is defined as the vector that maximizes the SNR,
\begin{equation}
    \mathbf{f} = \arg\max_{\mathbf{n}} R^2(\mathbf{n}), 
    \label{Eq:optimal_fingerprint}
\end{equation}
by employing Lagrange multiplier methods \cite{10_Methods2}, the optimal fingerprint $\mathbf{f}$ relates to the signal $\boldsymbol{\Psi}_{\rm s}$ as
\begin{equation}
    \mathbf{f} = \mathbb{C}^{-1} \boldsymbol{\Psi}^{\rm s},
    \label{Eq:optimal_fingerprint_solution}
\end{equation}
where $\mathbb{C}$ is the covariance matrix of the noise vector. 
Equation~\eqref{Eq:optimal_fingerprint_solution} shows that the optimal fingerprint $\mathbf{f}$ is, in general,
not parallel to the signal $\boldsymbol{\Psi}^{\rm s}$.
The optimal detection direction is neither along the direction of the maximum of the signal nor the minimum of the noise (see Fig.~\ref{fig:illustration_of_fingerprint}).

\begin{table}[t]
    \centering
    \caption{Mathematical elements required by OFM in climate system and spin magnetic resonance system}
    \label{tab:t1}
    \resizebox{\linewidth}{!}{
    \begin{tabular}{l@{\hskip 0.02\textwidth}p{0.25\textwidth}@{\hskip 0.02\textwidth}p{0.25\textwidth}}
        \hline\hline
        & \textbf{Climate System} \cite{10_Methods2,11_Methods1} & \textbf{Spin System}   \\
               \hline
        \textbf{External Forcing} & \textbf{Human and Natural Activities}, \newline e.g., Greenhouse gas emissions from human activities &  \textbf{Transverse External Magnetic Field}, \newline e.g., $B_1\cos\omega t$, $B_2\cos\omega t$ \\
        \hline
       \textbf{Model} & \textbf{Climate Models}, \newline e.g., Global Climate Models &  \textbf{Bloch Equation} \\
        \hline
        \textbf{Expected Signal} & \textbf{Response of climate variables to human activities according to climate model} & \textbf{Transverse component of spin as determined by the Bloch Equation}, \newline e.g., $\boldsymbol{\Psi}^{\rm s}$ \\
        \hline
        \textbf{Actual Signal} & \textbf{Measured data from various locations}, \newline e.g., Average surface temperature & \textbf{Output Signal in Experiment}, e.g., $\boldsymbol{\Psi}$  \\
        \hline
        \textbf{Noise} &  \textbf{natural variability} & \textbf{Noise in Spin Signal Induced by Magnetic Field Noise}, \newline e.g., $\delta \boldsymbol{\Psi}$  \\
        \hline\hline
    \end{tabular}}
\end{table}

Verifying the optimal fingerprint method requires four elements: (1) a set of observable variables; (2) governing equations that dictate the behavior of these variables; 
(3) control parameters driving changes in the variables; and (4) uncontrollable random noise. With these elements, optimal fingerprint method is essentially a technique 
for extracting information about changes in control parameters from data on observed variables that are contaminated by noise.

While OFM is typically applied within the framework described above, its direct verification in the actual climate is precluded by the system's inherent complexity and uncontrollability.
Instead, we emphasize that validation hinges on an analog system's ability to replicate the core mathematical structure, not its physical resemblance to the natural system. 
To this end, we select a minimal model—a spin resonance system—that enables a precise and controllable test of the method. 
Table 1 summarizes the mapping of mathematical elements between the two systems, and the application of OFM to the spin system is demonstrated in the following section.

\section{LINEAR RESPONSE THEORY AND GREEN'S FUNCTION}
\label{section_LRT}
Since Hasselmann's pioneering use of stochastic dynamics to model climate evolution, fluctuations arising from unresolved processes are represented as stochastic forcing. 
Within this framework and informed by recent advances \cite{7_SCMs1,lucarini_detecting_2024,santos_gutierrez_aspects_2022}, we describe the system using the Itô stochastic differential equation (SDE):
\begin{equation}
    d\mathbf{x} = \left(\mathbf{F}(\mathbf{x})+\epsilon_p g_p(t) \mathbf{G_p}(\mathbf{x}) \right)dt + \Sigma(\mathbf{x})d\mathbf{W}_t,
\end{equation}
where $\mathbf{F}$ denotes the smooth deterministic drift, $\mathbf{G}_p$ represents a pattern-based perturbation with amplitude modulated by $\epsilon_p \ll 1$ and a bounded scalar function $g_p(t)$. 
The term $\mathbf{W}_t$ is a $p$-dimensional Wiener process ($p \ge 1$), $\Sigma$ is the diffusion matrix, and $\mathbf{x}$ denotes the climate state.

Recent studies have demonstrated that the response theory can predict the system's response to a certain perturbation.
Building on this, Lucarini and Chekroun derived the leading-order response of nonequilibrium systems to pattern forcing within the framework of LRT \cite{lucarini_detecting_2024}.
The expected signal $\boldsymbol{\Psi}s$ induced by the forcing $G_p$ can be expressed via a suitable lagged correlation of observables in the unperturbed system, encapsulated by the Green's function:
\begin{equation}
    \Psi_{k}^{\rm{s}} = \epsilon_p \int_{-\infty}^{t} ds \mathcal{G}_{\mathbf{\Psi}_k}^{p}(t-s)g_p(s)
\end{equation}
This formulation introduces the causal structure of Green's functions into DA studies, offering a physically grounded interpretation of the optimal fingerprinting method.

Lucarini and Chekroun further applied this framework to perform DA in an energy balance model and a planet simulator, demonstrating the effectiveness of the approach.
Although LRT yields accurate attribution in both cases (with $\beta$ estimates close to unity), these examples remain numerical simulations within known models and have yet to be tested in real physical systems. 
Experimental validation in a controlled laboratory setting is therefore necessary.
To this end, one must first identify the external forcing applied to the physical system and the corresponding observables. 
The causal relationship between them can then be characterized via the Green's function derived from LRT. 
This relationship may be calibrated through controlled pattern forcing and precise measurements, or the Green's function itself may be estimated using alternative methods. 
In the following, we employ a spin resonance model to instantiate each element of the above framework, thereby assessing its validity and reliability in a real physical system.

\section{Fingerprint in spin system}
\label{section_OFMINSPIN}
In the spin resonance system, the dynamical model is fully described by the Bloch equation, 
with the external forcing being the transverse magnetic field components $B_x$ and $B_y$. 
The transverse spin components $\langle S_x\rangle$ and $ \langle S_y \rangle$ are the observed signals, 
and the spin noise induced by magnetic field fluctuations represents the noise in our observation(As shown in Table \ref{tab:t1}).
The Bloch equation is \cite{PhysRev.70.460,RevModPhys.44.169}
\begin{equation}
    \frac{d \mathbf{S}}{dt} = -\gamma \mathbf{B} \times \mathbf{S} - \hat{\boldsymbol{\Gamma}}  \cdot \mathbf{S}, 
    \label{Eq:Bloch_equations} 
\end{equation}
where $\mathbf{S}$ is the spin vector, $\gamma$ is the gyromagnetic ratio,
$\mathbf{B}$ is the magnetic field, 
and $\hat{\boldsymbol{\Gamma}} = \Gamma_2(\hat{\mathbf{x}}\hat{\mathbf{x}}+\hat{\mathbf{y}}\hat{\mathbf{y}}) + \Gamma_1 \hat{\mathbf{z}}\hat{\mathbf{z}}$ 
is the relaxation tensor with $\Gamma_1$ and $\Gamma_2$ being the longitudinal and transverse relaxation rate respectively \cite{PhysRevA.58.1412,franzen_spin_1959}.
The magnetic field $\mathbf{B}$ consists of a static field $B_z = B_0$ along the $\hat{\mathbf{z}}$ direction, 
which defines the Larmor frequency $\omega_0 = \gamma B_0$ of the spins,
and time-dependent fields $B_{x}(t)$ and $B_{y}(t)$ along $\hat{\mathbf{x}}$ and $\hat{\mathbf{y}}$ directions, which drive the precession of the spins.
To be specific, we consider the following form of the transverse driving field
\begin{eqnarray}
B_x(t) &=& \left[B_1(t) + \delta b(t) \right] \cos\omega t, \label{Eq:Bx}\\
B_y(t) &=& B_2(t) \cos\omega t.\label{Eq:By}
\end{eqnarray} 
The $B_1(t)$ and $B_2(t)$ act as the "pattern forcing" similar to the human activities.
Additionally, a Gaussian white noise term $\delta b(t)$ is superimposed on $B_x(t)$ to represent natural noise.
The correlation function of $\delta b(t)$ is $\langle \delta b(t)\delta b(t')\rangle =2D\delta_{\tau_{\rm c}}(t-t')$,
where $D$ characterizes the power spectrum density,
and $\delta_{\tau_{\rm c}}(t-t')$ is a Dirac-$\delta$ like function with short correlation time $\tau_{\rm c}$.

In the weak driving limit,  i.e., $B_0 \gg B_1$, $B_2$ and $\delta {b}(t)$, 
the $S_{z}$ could be treated as a constant, so the observed signal $\boldsymbol{\Psi}$ should be a two-dimensional vector consisting of transverse components of the spin, namely, 
\begin{equation}
    \boldsymbol{\Psi} \equiv 
    \begin{pmatrix}
    \tilde{S}_{x}\\
    \tilde{S}_{y}
    \end{pmatrix},
    \end{equation}
where $\tilde{S}_x = \Re\{(S_+e^{-i\omega t}\}$ and $\tilde{S}_y = \Im\{S_+e^{-i\omega t}\}$ denote the corresponding components in the rotating frame, and $S_{+} = S_x + i S_y$ is the spin component in the laboratory frame.
By solving the aforementioned stochastic differential equation, we derive the Green's function, which describes how the observed signal $\boldsymbol{\Psi}^{\rm s}$ responds to the "pattern forcing" $B_1$ and $B_2$.
\begin{equation}
    \label{Eq:Green'sFunction}
    \begin{split}
        \mathcal{G}_{\boldsymbol{\Psi}^{\rm s}}^{B_1}(t)  &= \Theta(t) {\gamma S_z}
        \begin{pmatrix}
            0 \\ 1
        \end{pmatrix}
        e^{-\Gamma_2 t},
        \\
        \mathcal{G}_{\boldsymbol{\Psi}^{\rm s}}^{B_2}(t) &= \Theta(t) {\gamma S_z}
        \begin{pmatrix}
            -1 \\ 0
        \end{pmatrix}
        e^{-\Gamma_2 t}.
    \end{split}
\end{equation}

\begin{figure}
    \centering \includegraphics{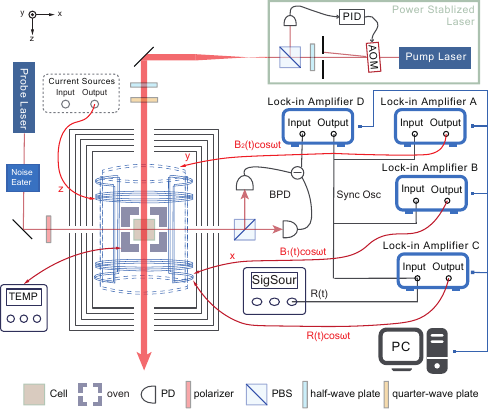} 
    \caption{\label{fig:-EXPsetup} Schematic illustration of the experiment setup. 
    Atomic spins are polarized and detected by the pump and probe laser beams.
    The optical signal is converted to an electrical signal by the balanced photodetector (BPD) and analysed and recorded by the data acquisition system consisting of a lock-in amplifier and a computer.
    The coil driver (including four synchronized lock-in amplifiers and a signal source) generates electric current with well controlled amplitude and phase to drive the transverse magnetic field $(B_1 \hat{\mathbf{x}}+B_2 \hat{\mathbf{y}})\cos\omega t $
    and the Gaussian noise field $b(t)\cos\omega t$ (see Appendix \ref{section_experiment_ap} for experimental details).}
\end{figure}

It is worth mentioning that weak driving condition also guarantees the prerequisite of the optimal fingerprint method, namely, 
the expected signal $\boldsymbol{\Psi}^{\rm s}$ can be separated from the noise signal $\delta \boldsymbol{\Psi}$ driven by $\delta b(t)$. 
In the resonant case, one obtains the expected signal $\boldsymbol{\Psi}^{\rm s}$ and the covariance matrix of $\delta\boldsymbol{\Psi}$ as

\begin{eqnarray}
    \boldsymbol{\Psi}^{\rm s} = 
    \frac{\gamma S_{z}}{2 \Gamma_2}
    \begin{pmatrix}
        -B_{2} \\
        B_{1} 
    \end{pmatrix}.
\end{eqnarray}

\begin{equation}\label{eq:covariance matrix with resonance}
    \mathbb{C} = \frac{D\gamma^2 S_z^2}{8\Gamma_2} \begin{pmatrix}
        1 & 0 \\
        0 & 3
    \end{pmatrix}.
\end{equation}
According to Eqs.~\eqref{Eq:optimal_fingerprint_solution} \& \eqref{eq:covariance matrix with resonance}, the optimal detection direction $\theta_{\rm opt}$ depends on the driving field amplitudes $B_1$ and $B_2$ as
\begin{equation}
    \label{eq:fingerprint in spin system}
    \tan\theta_{\rm opt} = -\frac{B_1}{3B_2},
\end{equation}
which indicates the optimal fingerprint detection in the spin system. The detail theoretical  treatment is presented in Appendix \ref{section_theory_ap}.

The spin magnetic resonance system described by Eqs.~\eqref{Eq:Bloch_equations}-\eqref{eq:covariance matrix with resonance} 
supports a fully controllable physical model to experimentally study the optimal fingerprint method.
Various parameters in the spin system, including the relaxation rates $\Gamma_1$ and $\Gamma_2$, the noise intensity $D$, and the driving field amplitudes $B_1$ and $B_2$, are tunable in a wide range to mimic the climate system.
Furthermore, the resonance system contains several time scales, including
(i) the correlation time $\tau_{\rm c}$ of the noise $\delta b(t)$ is the shortest timescale in the system;
(ii) the spin precession period $T_0 = 2\pi/\omega_0$;
(iii) the spin relaxation time $T_1 = \Gamma_1^{-1}$ and $T_2 = \Gamma_2^{-1}$  \cite{Seltzer2008DevelopmentsIA,RevModPhys.44.169,PhysRevA.58.1412,franzen_spin_1959}; and
(iv) the characteristic timescale $\tau_{\rm B}$ of the change of the driving field $B_{1,2}(t)$, which changes very slowly and is the longest timescale in the system.
The hierarchy of different time scales (i.e. $\tau_{\rm c} \ll T_0 \ll T_{1,2} \ll \tau_{\rm B}$) mimics the complex dynamic behavior of climate system. 
Consequently, the spin resonance model is a suitable physical system for verifying fingerprint method.

\begin{figure}
    \centering
    \includegraphics[scale=0.60]{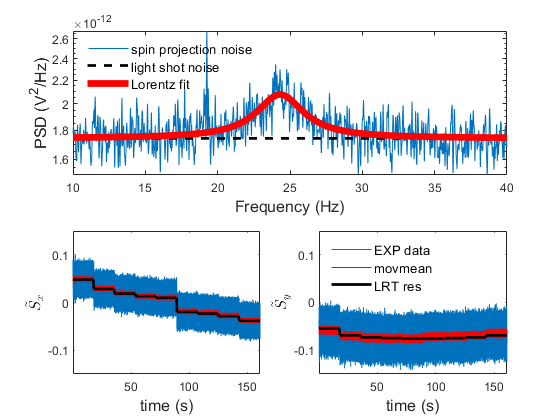}
    \caption{\label{fig:-SpinProjectionNoiseAndGreenFunction}
            Spin projection noise and signal attribution. (a) The noise power spectrum density demonstrates the spin projection noise under no external drive($B_x = B_y = 0$).
            According to the fluctuation-dissipation theorem(see Appendix\ref{section_GreenFunction_ap}), the PSD result is consistent with the form of the Green's function given by Eq.(\ref{Eq:Green'sFunction}).
            (b) and (c) display the expected signals $\boldsymbol{\Psi}^{s}$ varying with "pattern forcing"  $B_1(t)$ and $B_2(t)$ and natural noise $\delta b(t)$.
            The measured $\tilde{S}_x$ and $\tilde{S}_y$ are shown in blue, with their moving average overlaid in red.            
            The black line, representing the theoretical prediction from linear response theory, is in excellent agreement with this experimental average.
    }
\end{figure}

\begin{figure}
    \centering
    \includegraphics[scale=0.65]{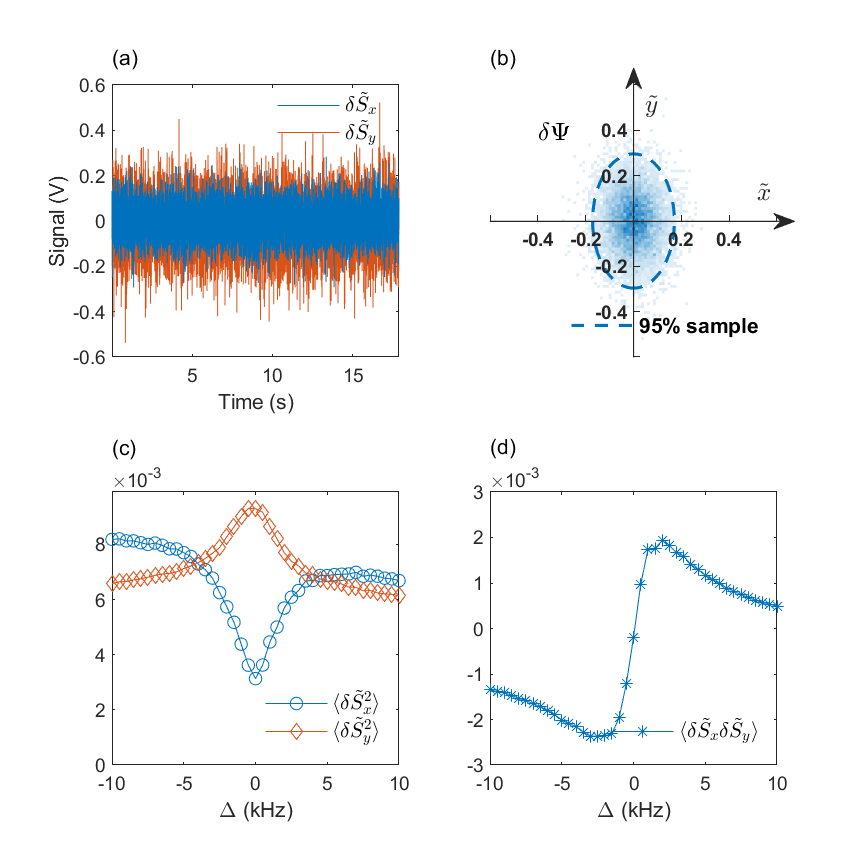}
    \caption{\label{fig:-Timedomain and covariance}
        (a) Measured spin time-domain fluctuation $\delta \tilde{S}_x$ and $\delta \tilde{S}_y$ in the rotating frame. 
        (b) Distribution of the spin fluctuation represented by the vector $\delta\boldsymbol{\Psi}$ in the two-dimensional plane. 
        With the resonance condition ($\Delta = 0$), the distribution is an ellipse with the major axis along $\tilde{S}_y$ direction, and the aspect ratio is $1/3$, which agrees with the prediction in Eq. (\ref{eq:covariance matrix with resonance}). 
        (c) \& (d) The variances $\langle\delta\tilde{S}_x^2\rangle$ and $\langle\delta\tilde{S}_y^2\rangle$ [the diagonal elements of covariance matrix] 
        and the cross correlation function $\langle\delta\tilde{S}_x\delta\tilde{S}_y\rangle$ [the off-diagonal element of covariance matrix] as functions of the detuning $\Delta$. 
        The detailed calculation can be found in the supplementary. 
        } 
\end{figure}

\section{Experimental verification of fingerprint method}
\label{section_experiment}
The experiment setup is illustrated in Fig.~\ref{fig:-EXPsetup}. 
A glass cell filled with purified $^{87}\mathrm{Rb}$ atoms and 350 Torr $N_2$ as buffer gas was placed in an oven, which is heated to $110^{\circ}{\rm C}$. 
The atomic density of the Rb vapor at this temperature is $1.2\times 10^{13}~{\rm cm}^{-3}$. 
The glass vapor cell, the oven and a set of three-axis magnetic field coils are placed in a five-layer magnetic shield to eliminate the earth magnetic field and the unwanted magnetic noise.
The Rb atomic spins are optically pumped by a circular polarized laser beam propagating along the $\hat{\mathbf{z}}$ direction in resonance with the D1 line of the Rb transition.
The transverse spin component $S_x$ of Rb atoms in lab coordinates is detected by a linearly polarized laser beam along $\hat{\mathbf{x}}$ direction via the Faraday rotation effect  \cite{Seltzer2008DevelopmentsIA}. 
A magnetic field $B_0 = 3.5~\mathrm{\mu T}$ is applied along $\hat{\mathbf{z}}$ direction, corresponding to the Larmor frequency $\omega_0 = 24.75~\mathrm{kHz}$ of $^{87}{\rm Rb}$ spins.
The deterministic amplitudes $B_1$ and $B_2$ of the driving field components are generated by the LIA, and the Gaussian noise $\delta b(t)$ is generated by a signal source. 
These signals are converted to the driving magnetic fields by the coil driver (see Appendix \ref{section_experiment_ap} for experimental details).

Experimentally, the signal $\boldsymbol{\Psi}$ is obtained directly by demodulating the voltage $V(t)$ from the balanced photodetector (i.e. BPD, see Fig.~\ref{fig:-EXPsetup}) 
with a reference signal $V_{\mathrm{ref}}(t) = \sqrt{2}e^{-i(\omega t+\theta)}$ via a lock-in amplifier (LIA) , where $\theta$ is the demodulation phase  \cite{Meade1983LockinA}.
Given the phase $\theta$, the demodulation output 
\begin{equation}
V_{\mathrm{out}}(\theta) = \boldsymbol{\Psi} \cdot \mathbf{n}(\theta)
\end{equation}
is equivalent to a projection of the signal vector $\boldsymbol{\Psi}$ along the direction $\mathbf{n}(\theta) = (\cos\theta,\sin\theta)$ in the rotating frame.
The fingerprint method is examined by measuring the SNR of the output $V_{\mathrm{out}}(\theta)$ of the spin system as a function of $\theta$. 
The optimal fingerprint $\mathbf{f}$ in Eq.~(\ref{Eq:optimal_fingerprint_solution}) is indeed the projection direction $\mathbf{n}(\theta)$ which maximizes the SNR.

\begin{figure}
    \centering
    \includegraphics[scale=0.65]{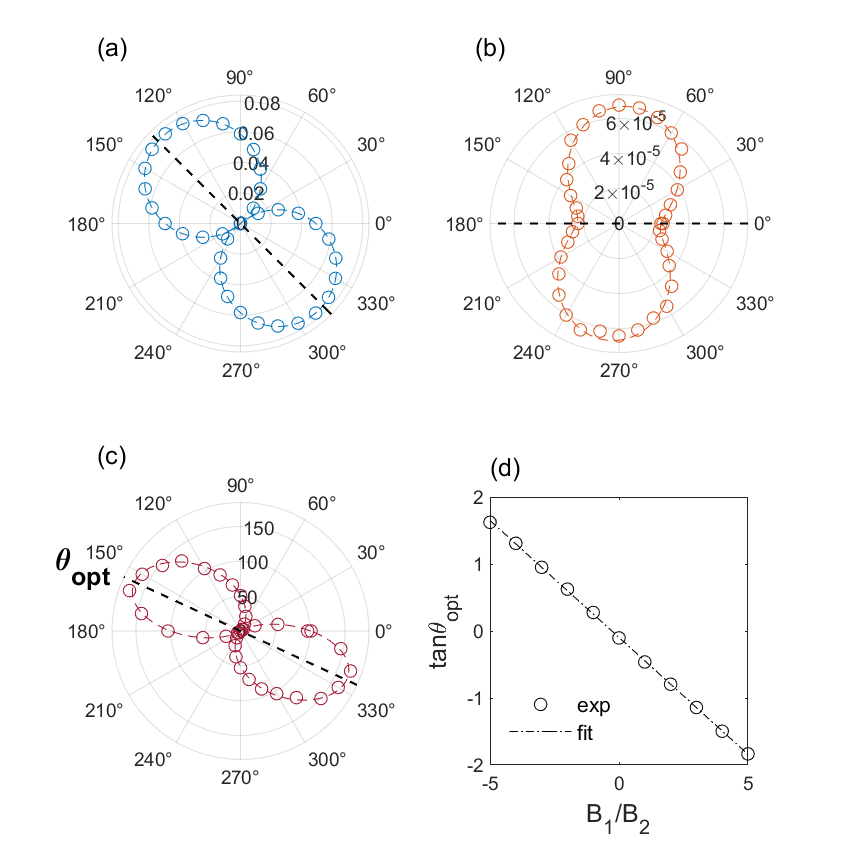}
    \caption{\label{fig:-SNR distribution}
        (a) The spin signal $\langle\boldsymbol{\Psi}^{\rm s}\cdot\mathbf{n}(\theta)\rangle$ as a function of the detection direction $\mathbf{n}(\theta)$.
        With amplitude ratio of the driving field set to $B_1/B_2 = 1$, the detection direction which maximized the signal amplitude should be $\theta = 135^\circ$ as shown by the dashed line.
        (b) The noise amplitude $\langle(\delta\boldsymbol{\Psi}_s\cdot\mathbf{n}(\theta))^2\rangle$ as a function of the detection direction $\mathbf{n}(\theta)$.
        The noise amplitude is minimized when the detection is along the $\tilde{x}$ axis, the dashed line with $\theta = 0^\circ$.
        (c) The SNR as a function of the detection direction $\mathbf{n}(\theta)$. 
        The measured data shows the optimal detection direction which maximizes the SNR at $\theta_{\rm opt} \approx 155^\circ$, which is in good agreement with the theoretical prediction of Eq.~\eqref{eq:fingerprint in spin system} as shown by the dashed line.
        (d) The measured optimal fingerprint (the symbols) as a function of the ratio $B_1/B_2$. The dash-dotted line is the theoretical prediction according to Eq.~(\ref{eq:fingerprint in spin system})}
\end{figure}

We first examine LRT in the spin system. 
To this end, we estimated the Green's function from spin-projection noise by using the fluctuation-dissipation theorem. in the absence of external driving fields (i.e., $B_1 = B_2 = 0$, $\delta b(t) = 0$).
According to the fluctuation-dissipation theorem, the linear response of a system—characterized by its Green's function—is encoded in its spontaneous fluctuations, which in this case correspond to spin projection noise.
In the undriven regime, the power spectral density of spin projection noise $\mathcal{S}_{\mathrm{SPN}}$, arising from intrinsic spin fluctuations, is given by Eq. (\ref{Eq:spinnoise}) (see Appendix \ref{section_GreenFunction_ap} for a detailed derivation):
\begin{equation}
    \label{Eq:spinnoise}
    \mathcal{S}_{\rm SPN} = \frac{\gamma S_z}{\Gamma_2} \cdot \frac{1}{\Gamma_2^2 + \Delta ^2}
\end{equation}
The experimentally measured spin projection noise, shown in Fig. \ref{fig:-SpinProjectionNoiseAndGreenFunction}(a), is in good agreement with Eq. (\ref{Eq:spinnoise}). 
Having established the Green's function from these measurements, we then apply it to attribute the sinusoidal driving fields $B_1$ and $B_2$. 
The results, presented in Fig. \ref{fig:-SpinProjectionNoiseAndGreenFunction}(b), accurately capture the influence of $B_1(t)$ and $B_2(t)$ on the observed signal $\boldsymbol{\Psi}^{\rm s}$.

We then use OFM to separate the expected signal $\boldsymbol{\Psi}^{\rm s}$ from the natural noise.
Figure.~\ref{fig:-Timedomain and covariance} shows the measured spin signal vector $\boldsymbol{\Psi}$ and its statistical properties under the driving fields $B_x(t)$ and $B_y(t)$ in Eqs.~\eqref{Eq:Bx} \& \eqref{Eq:By}.
In the case of resonant driving ($\Delta = 0$), with the demodulation phase $\theta = 0$ or $\theta= \pi/2$, the fluctuation of $\tilde{S}_x(t)$ and $\tilde{S}_y(t)$ induced by the noise field $\delta b(t)$ is obtained, as shown in Fig.~\ref{fig:-Timedomain and covariance}(a).
In the two-dimensional plane spanned by $\tilde{S}_x(t)$ and $\tilde{S}_y(t)$, the distribution of the spin signal vector $\boldsymbol{\Psi}$ forms an ellipsoid with an aspect ratio of $\langle\delta\tilde{S}_y^2\rangle/\langle\delta\tilde{S}_x^2\rangle = 3$ [see Fig.~\ref{fig:-Timedomain and covariance}(b)],
which agrees with the theoretical prediction of Eq.~\eqref{eq:covariance matrix with resonance}.
The full covariance matrix $\mathbb{C}$ of the spin fluctuation $\delta\boldsymbol{\Psi}$ and its frequency dependence is also measured by varying the driving frequency $\omega$.
The variance of $\delta\tilde{S}_y$ ($\delta\tilde{S}_x$) shows a Lorentzian (anti-Lorentzian) lineshape, and the correlation function $\langle\delta\tilde{S}_x\delta\tilde{S}_y\rangle$ is of dispersive profile [Figs.~\ref{fig:-Timedomain and covariance}(c) \& (d)].

Measuring the signal along the $\tilde{x}$ or $\tilde{y}$ axes is not the optimal choice to extract the signal from the noise. 
Although the variance of the signal $\langle\tilde{S}_x^2\rangle$ is small, the signal amplitude along the $\tilde{x}$ axis $\langle\tilde{S}_x\rangle$ is also small.
By setting the demodulation phase $\theta$, we choose a measurement direction $\mathbf{n}(\theta)$, along which the signal amplitude is a linear combination of $\langle\tilde{S}_x\rangle$ and $\langle\tilde{S}_y\rangle$
\begin{equation}\label{eq:EXPSNRdefine_1}
        \langle \boldsymbol{\Psi}_s\cdot \mathbf{n}(\theta)\rangle =  \cos\theta\langle\tilde{S}_x\rangle + \sin\theta \langle\tilde{S}_y\rangle,
\end{equation}
and under resonance driving condition, the correlation function $\langle\delta\tilde{S}_x\delta\tilde{S}_y\rangle = 0$. The variance is 
\begin{equation}\label{eq:EXPSNRdefine_2}
        \langle (\delta\boldsymbol{\Psi}\cdot \mathbf{n})^2 \rangle =\cos^2\theta \langle\delta\tilde{S}^2_x\rangle +\sin^2\theta \langle\delta\tilde{S}^2_y\rangle.
\end{equation}
To find the maximum SNR, we record the signal amplitude and the variance simultaneously while sweeping the measurement direction $\mathbf{n}(\theta)$.
As shown in Figs.~\ref{fig:-SNR distribution}(a) \& \ref{fig:-SNR distribution}(b),
both signal amplitude and variance have their own maximum and minimum values, but they do not coincide in the same direction $\mathbf{n}(\theta)$. 
According to the fingerprint method, there exists an optimal detection direction $\mathbf{n}(\theta_{\rm opt})$ which maximizes the SNR. 
Figure.~\ref{fig:-SNR distribution}(c) shows the SNR as a function of the detection angle $\theta$, and the optimal detection direction $\mathbf{n}(\theta_{\rm opt})$ is indicated by the black dashed line.

Figure.~\ref{fig:-SNR distribution}(d) shows the measured the optimal detection direction $\theta_{\rm opt}$ as a function of the ratio $B_1/B_2$,
and the dashed line indicates the theoretical prediction according to Eq.\eqref{eq:fingerprint in spin system}.
The linear dependence between experiment data and theory agrees well indicating that the fingerprint method in the spin resonance system has been verified.

\section{Conclusion}
\label{section_conclusion}
The optimal fingerprint method (OFM) has played a pivotal role in advancing modern climate science, profoundly shaping both scientific understanding and policy decisions. 
Concurrently, the development of stochastic dynamical models and LRT has provided a physically grounded framework for DA studies. 
In this work, we experimentally validate two core aspects of DA—the linear response model and the signal-to-noise optimization underlying OFM—using a controlled spin system as a physical analog. 
Under the weak driving approximation, the spin system can be well linearized.
By leveraging the fluctuation-dissipation theorem, we measured the system's Green's function via spin projection noise and used its causal structure to attribute spin signals to external driving fields, 
thereby confirming the theoretical predictions of Lucarini et al \cite{lucarini_detecting_2024}. 
Furthermore, we applied Hasselmann's original OFM concept to separate the expected signal from system noise. 
Through statistical characterization of the noise, we identified the optimal measurement direction that maximizes the signal-to-noise ratio and verified this prediction experimentally. 
Our results provide experimental support for the application of OFM and LRT in real physical systems, 
and underscoring their application potential in ecosystem, finance, social sciences, quantum sensing and other interdisciplinary fields.

In addition, Lucarini et al also analyze the theoretical basis and application value of the nonlinear response of OFM in the nonlinear system. 
The follow-up work can further carry out experimental demonstration in this direction. Under the condition of breaking the weak driving approximation, 
the spin system can also reflect the nonlinear behavior due to the spin exchange interaction, which also provides a potential physical application scenario for the subsequent verification of the nonlinear response theory of OFM.

\begin{acknowledgments}
This work is supported by NSFC (Grants No. 12088101,  No. U2030209 and No. U2230402).
\end{acknowledgments}

\appendix
\section{THEORETICAL ANALYSIS OF FINGERPRINT METHOD IN SPIN SYSTEM}

\label{section_theory_ap}
The Bloch equation is presented as
\begin{equation}
    \frac{d \mathbf{S}}{dt} = -\gamma \mathbf{B} \times \mathbf{S} - \hat{\boldsymbol{\Gamma}}  \cdot \mathbf{S} + \mathbf{R}, 
    \label{Eq:Bloch_equations_ap} 
\end{equation}
where $\mathbf{S}$ is the spin vector of Rb atoms, $\gamma$ is the gyromagnetic ratio, and 
$\hat{\boldsymbol{\Gamma}} = \Gamma_2(\hat{\boldsymbol{x}}\hat{\boldsymbol{x}}+\hat{\boldsymbol{y}}\hat{\boldsymbol{y}})+\Gamma_1\hat{\boldsymbol{z}}\hat{\boldsymbol{z}}$ 
is the relaxation tensor with $\Gamma_1$ and $\Gamma_2$ being the longitudinal and transverse relaxation rate respectively.
We consider the situation that the pump light is along the $\hat{\boldsymbol{z}}$ direction, so the pumping term in Eq.~\eqref{Eq:Bloch_equations_ap} is $\mathbf{R} = R_0\hat{\boldsymbol{z}}$ 
where $R_0$ is the pumping rate.
The magnetic field $\mathbf{B} =  B_x\hat{\boldsymbol{x}}+B_y\hat{\boldsymbol{y}}+B_z\hat{\boldsymbol{z}}$ is applied in the following form
\begin{equation}
    \begin{split}
        B_x &= B_1(t) \cos\omega t + \delta b(t)  \cos(\omega t+\phi ), \\
        B_y &= B_2(t) \cos\omega t,\\
        B_z &= B_0.\label{Eq:Magnetic_field}    
    \end{split}
\end{equation} 
A large static magnetic field $B_0$ is settled along the $\hat{\boldsymbol{z}}$ direction parallel to the pumping light as a main magnetic field, defining the Larmor frequency $\omega_0 = \gamma B_0$.
A transverse driving field $\tilde{B}_+(t)\cos\omega t =  (B_1(t)+iB_2(t))\cos\omega t$ is settled to maintain spin precession. 
The angle $\psi$ between the direction of the driving field and the $\hat{\boldsymbol{x}}$ axis satisfies $\tan\psi = B_2/B_1$.
Furthermore, a modulated Gaussian white noise $\delta b(t)\cos(\omega t+\phi)$ is applied along the $\hat{\boldsymbol{x}}$ direction,
where $\delta b(t)$ is a Gaussian white noise with autocorrelation function
$\langle\delta b(t_1)\delta b(t_2)\rangle = 2D\delta(t_1-t_2)$, and $2D$ is the variance of white noise.
The angle $\phi$ is the phase delay caused by the modulation process, we will discuss its effect in the later part of this supplementary material.

To study the precession signal, we define the complex transverse spin component $S_+ = S_x + iS_y$, whose equation of motion is
\begin{equation}
     \frac{dS_+}{dt} = -i\omega_0 S_+ - \Gamma_2 S_+ +i\gamma S_z \left[\tilde{B}_+\cos \omega t +\delta b(t)\cos( \omega t+\phi)\right].
\end{equation}

Solving the Bloch equation in the rotating frame is a mature and convenient method. 
In the rotating frame, the transverse spin signal becomes $\tilde{S}_+ = S_+ e^{i\omega t}$, and the equation is expressed as
\begin{equation}
    \frac{d \tilde{S}_+}{dt} = -( i\Delta +\Gamma_2) \tilde{S}_+ + i\gamma S_z \left[\tilde{B}_{+} \cos\omega t + \delta b(t)\cos(\omega t + \phi)\right] e^{i\omega t},
    \label{eq:Bloch_equations_in_rotating_frame}
\end{equation}
where $\Delta = \omega_0 - \omega$ is the detuning. 
In the weak driving limit ($B_1,B_2,\delta b \ll B_0$), the longitudinal component $S_z$ is constant
\begin{equation}
    S_z = \frac{R_0}{\Gamma_1}\frac{\Delta^2 +\Gamma_2^2}{\Delta^2 +\Gamma_2^2(1+s)},    
\end{equation}
where $s = \gamma^2(B_1^2+B_2^2)/\Gamma_1\Gamma_2$ is the saturation factor. 
The solution of Eq.~(\ref{eq:Bloch_equations_in_rotating_frame}) in this case is
\begin{widetext}
    \begin{equation}
        \tilde{S}_{+} = i\gamma S_{z}e^{-(i\Delta+\Gamma_{2})t}\bigg( 
        \int_{0}^t  \, dt'  \tilde{B}_+\cos \omega t'e^{(i\omega_{0}+\Gamma_{2})t'}
        +\int_{0}^t   \, dt'  \delta  b(t') \cos (\omega t'+\phi)e^{(i\omega_{0}+\Gamma_{2})t'}
        \bigg  ) \equiv  \langle{\tilde{S}_{+}}\rangle + \delta \tilde{S}_{+}.
    \label{eq:solution_of_Blocheq}
    \end{equation}
\end{widetext}

Here, $\langle{\tilde{S}_{+}}\rangle$ and $\delta \tilde{S}_{+}$ indicate the average part (the signal) and fluctuating (the noise) of transverse spin component.

Under the rotating wave approximation (RWA) and in the long-time limit (i.e., $t\gg 1/\Gamma_2$), the average part $\langle{\tilde{S}_{+}}\rangle$ is

\begin{equation}
    \begin{split}
        \langle{\tilde{S}_{+}}\rangle = \frac{S_z}{2}\frac{-\omega_2\Gamma_2+\omega_1\Delta + i(\omega_1\Gamma_2+\omega_2\Delta)}{\Gamma^2+\Delta^2},
    \end{split}
\end{equation}
and the signal vector is
\begin{eqnarray}
    \begin{split}
        \boldsymbol{\Psi}^{\rm s} &= 
        \begin{pmatrix}
        \langle\tilde{S}_{x}\rangle\\     
        \langle\tilde{S}_{y}\rangle
        \end{pmatrix} 
        =
        \begin{pmatrix}
        {\rm Re}[\langle\tilde{S}_{+}\rangle] \\
        {\rm Im}[\langle\tilde{S}_{+}\rangle] 
        \end{pmatrix} \\
        &= 
        \frac{\gamma S_{z}}{2(\Delta^2 + \Gamma_2^2)}
        \begin{pmatrix}
            B_{1}\Delta -  B_{2} \Gamma_2\\
            B_{1} \Gamma_2 +  B_{2}\Delta
        \end{pmatrix},
        \label{eq:Signal_in_fingerprint}
    \end{split}
\end{eqnarray}
which is determined by the driving field amplitude $B_1$, $B_2$ and driving detuning $\Delta$.


As shown in the Eq.~(\ref{eq:solution_of_Blocheq}), the modulated Gaussian noise $b(t)\cos\omega t$  induces the fluctuation $\delta \tilde{S}_+(t)$ in the spin signal 
\begin{equation}
    \delta \tilde{S}_+(t) = i\gamma S_{z}e^{-(i\Delta+\Gamma_2)t}\int_{0}^{t}dt'\delta  b(t') \cos (\omega t'+\phi)e^{(i\omega_{0}+\Gamma_{2})t'}.
\end{equation}

Here we consider an additional modulation phase $\phi$ caused by the imperfect hardware implementation and try to analyze how it affects the measurement results. 
The source of this additional phase and how to control it will be discussed in the later section. In this case, the noise covariance matrix is

\begin{widetext}
    \begin{eqnarray}\label{equ:covariance}
        \mathbb{C} = \left(
        \begin{array}{cc}
            \langle\delta S_x^{'2}\rangle & \langle\delta S_x\delta S_y \rangle  \\
            \langle\delta S_x \delta S_y\rangle  & \langle\delta S_y^{'2}\rangle 
        \end{array}
        \right) 
        = \frac{D\gamma^2 S_z^2}{8(\Delta^2 + \Gamma_2^2)} \begin{pmatrix}
            -\Gamma_2\cos2\phi - \Delta \sin2\phi & \Gamma_2\sin2\phi + \Delta \cos2\phi  \\
            \Gamma_2\sin2\phi + \Delta \cos2\phi  & \Gamma_2\cos2\phi + \Delta \sin2\phi
        \end{pmatrix} + \frac{D\gamma^2 S_z^2}{4 \Gamma_2} \mathbb{I}.
    \end{eqnarray}        
\end{widetext}
To focus on the impact of detuning to the covariance matrix,  we consider the covariance matrix with phase $\phi = 0$, i.e.,
\begin{eqnarray}\label{eq:covariance matrix}
    \mathbb{C}
    &=& \frac{D\gamma^2 S_z^2}{8(\Delta^2 + \Gamma_2^2)} \begin{pmatrix}
        -\Gamma_2 & \Delta \\
        \Delta  & \Gamma_2
    \end{pmatrix} + \frac{D\gamma^2 S_z^2}{4 \Gamma_2} \mathbb{I}. 
    \label{eq:covariance matrix general}
\end{eqnarray}
It's necessary to discuss how the noise power distribution in the frequency domain to choose an appropriate experimental measurement bandwidth including all the noise information.
By treating the modulated Gaussian noise as a cyclostationary process signal, we can calculate the average power spectral density (PSD) of $ S_x$ and $ S_y$ by performing a Fourier transform 
on the zero cyclic frequency autocorrelation 
\begin{widetext}
    \begin{eqnarray}\label{equ:spinPSD}
        \begin{split}
            G_{ \delta S_x'}(\omega)=&-\frac{1}{8} \frac{D\gamma^2 S_{z0}^2}{\Delta^2+\Gamma_2^2} \frac{\Gamma^2-(\omega-\Delta)\Delta}{(\omega-\Delta)^2+\Gamma_2^2} + \frac{D\gamma^2 S_{z0}^2}{4}\frac{1}{(\omega-\Delta)^2+\Gamma_2^2}
            \\
            G_{ \delta S_y'}(\omega)=&\frac{1}{8} \frac{D\gamma^2 S_{z0}^2}{\Delta^2+\Gamma_2^2} \frac{\Gamma^2-(\omega-\Delta)\Delta}{(\omega-\Delta)^2+\Gamma_2^2} + \frac{D\gamma^2 S_{z0}^2}{4}\frac{1}{(\omega-\Delta)^2+\Gamma_2^2}
        \end{split}
    \end{eqnarray}            
\end{widetext}

In most cases, the frequency of the driving field is set in resonant with the spins (i.e., $\Delta = 0$). 
So the PSD becomes
\begin{eqnarray}\label{equ:spinResPSD}
    \begin{split}
        G_{ \delta S_x'}(\omega)=& \frac{1}{8} \frac{D\gamma^2 S_{z0}^2}{\omega^2+\Gamma_2^2} ,
        \\
        G_{ \delta S_y'}(\omega)=&  \frac{3}{8} \frac{D\gamma^2 S_{z0}^2}{\omega^2+\Gamma_2^2} .
    \end{split}
\end{eqnarray}
Under the resonance condition, the PSD of $\delta\tilde{S}_x$ and $\delta\tilde{S}_y$ become to the Lorentzian profile at zero frequency with a linewidth of $\Gamma_2$. 
By setting the measurement bandwidth larger than $\Gamma_2$, We can directly extract complete variance information from time-domain signals. 
If $\Delta \neq 0$, we can consider the PSD approximately as a Lorentzian profile with center at $\Delta$, so in this case we need to set the measurement bandwidth larger than $\Gamma_2+\Delta$ to get all the variance information.

\begin{figure}
    \centering \includegraphics[scale=0.33]{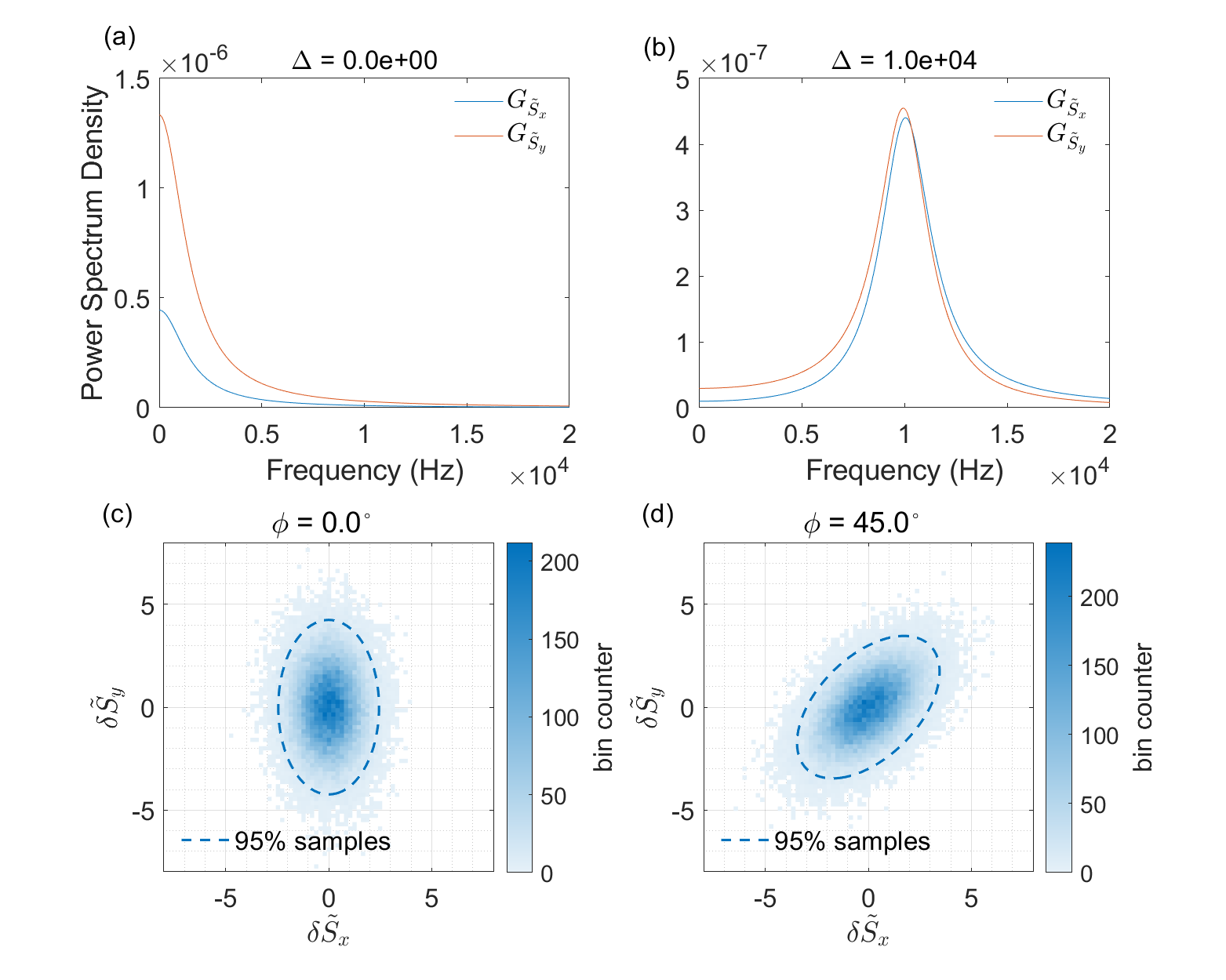} 
    \caption{\label{fig:noisedistrubution} 
    Figure (a) and (b) show the theoretical simulation of the PSD of $\tilde{S}_x$ and $\tilde{S}_y$ with different values of detuning $\Delta$.
    The noise distribution can be approximated as a Lorentzian lineshape with the center frequency $\Delta$ and the linewidth $\Gamma_2$.
    To obtain all the spin noise information, the detuning $\Delta$ need to be settled at zero and filter bandwidth needs to be larger than linewidth $\Gamma_2$.
    Figure (c) and (d) show the theoretical simulation of the noise distribution of $\delta \boldsymbol{\Psi}$ with different $\phi$. According to Eq.~(\ref{eq:covariance_withphi}), 
    different $\phi$ will cause different noise distribution, leading to different fingerprint results}
\end{figure}

In this section, we will focus on how the modulated phase delay $\phi$ affects the noise distribution and the fingerprint in the spin resonance system.
By controlling resonance driving $\Delta = 0$, the covariance matrix becomes
\begin{eqnarray}
    \mathbb{C} = \frac{D\gamma^2 S_z^2}{8\Gamma_2} \begin{pmatrix}
        2 - \cos2\phi  & \sin2\phi   \\
        \sin2\phi  & 2 + \cos2\phi 
    \end{pmatrix}, 
    \label{eq:covariance_withphi}
\end{eqnarray}
The eigenvalue ${\Sigma}^2$ and normalized eigenvector $\mathbb{V}$ of Eq.~(\ref{eq:covariance_withphi}) is
\begin{equation}
    \begin{split}
        {\Sigma}^2 &=
        \begin{pmatrix}
            \sigma_1^2 & 0   \\
            0  & \sigma_2^2 
        \end{pmatrix}
        = 
        \begin{pmatrix}
            3 & 0   \\
            0  & 1 
        \end{pmatrix},\\ 
        \mathbb{V} &= 
        (\mathbf{v}_1.\mathbf{v}_2)
        =
        \begin{pmatrix}
            \sin\phi  & -\cos\phi   \\
            \cos\phi  & \sin\phi 
        \end{pmatrix}.
    \end{split}
\end{equation}

The covariance matrix satisfies $\mathbb{C} = \mathbb{V}{\Sigma}^2\mathbb{V}^\mathsf{T}$.
By comparing the covariance matrix with the quadratic equation of an ellipse  
\begin{equation}
    \delta \boldsymbol{\Psi}^\mathsf{T}\mathbb{C}^{-1}\delta \boldsymbol{\Psi} = \delta \boldsymbol{\Psi}^\mathsf{T} \mathbb{V} (\Sigma^2)^{-1} \mathbb{V}^{-1} \delta \boldsymbol{\Psi} = s, 
\end{equation}
we can get the noise distribution in the rotating frame.
From the geometric perspective, the noise $\delta \boldsymbol{\Psi} = (\delta \tilde{S}_x,\delta \tilde{S}_y)^\mathsf{T}$ distributes in a ellipse with 

the major and minor axes in the direction of $\mathbf{v}_1 = (\sin\phi,\cos\phi)^\mathsf{T}$ and $\mathbf{v}_2 = (-\cos\phi,\sin\phi)^\mathsf{T}$ 
with length of $\sigma_1$ and $\sigma_2$.
The $s$ is the value of the chi-squared distribution $\chi^2_k$. In our situation the degrees of freedom $k = 2$, so that when choosing $s = 5.99$, 
the ellipse equation presents the area with $95\%$ probability of the noise distribution. The simulation result is displayed in Fig.~\ref{fig:noisedistrubution}, 
a noise realization containing fifty thousand samples described by Eq.~(\ref{eq:covariance_withphi}) is displayed, and the colorbar indicates the sample distribution number 
in every counter cells. Besides that the ellipse equation is also displayed by the dashed line, which can be confirmed to contain $95\%$ of the samples.

By changing the modulated phase delay, the major and minor axes of the ellipse will clockwise rotate following the phase delay $\phi$.
It seems to be easy to understand how the phase affects the noise distribution and to control the noise to meet our requirements. 
However, in the real experimental settings, owing to the modulated frequency choosing and the phase-frequency response of instruments, 
the modulated phase delay $\phi$ may be unpredictable and unstable, which will cause an incorrect and non repeatable noise measurement.
Appendix~\ref{section_experiment_ap} shows how the control of $\phi$ allows a stable measurement noise statistics.

According to fingerprint method, the optimal fingerprint relates to the signal $\boldsymbol{\Psi}^{\rm s}$ as
\begin{equation}
    \mathbf{f} = \mathbb{C}^{-1} \boldsymbol{\Psi}^{\rm s}.
    \label{Eq:optimal_fingerprint_solution_AP}
\end{equation}

\begin{figure}
    \centering \includegraphics[scale=0.60]{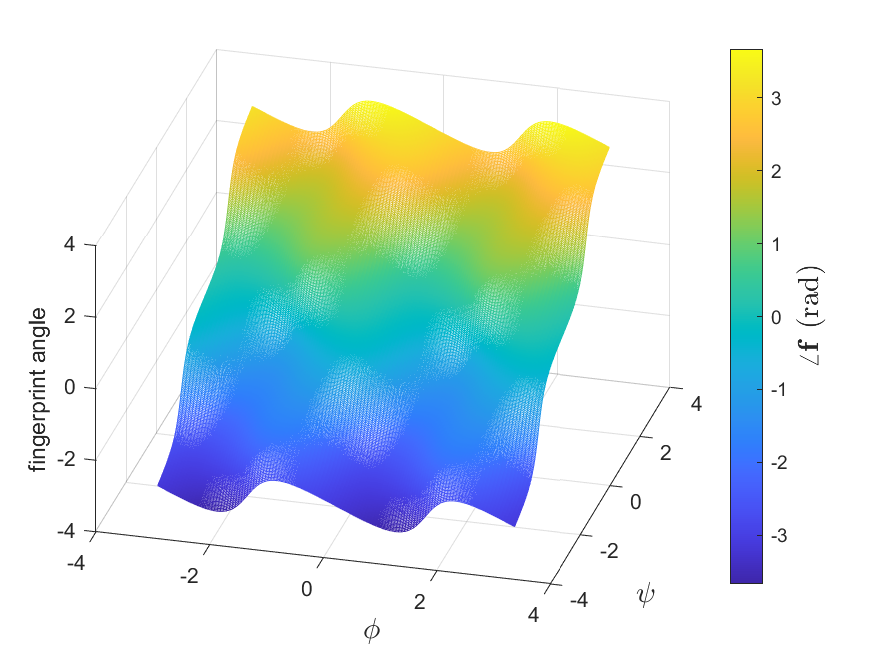} 
    \caption{\label{fig:fingerprintangle2} The dependence of angle of the fingerprint vector $\mathbf{f}$ on the driving direction $\psi$ and modulated phase delay $\phi$.}
\end{figure}

So the fingerprint direction under resonance condition is
\begin{equation}
    \begin{split}
        \tan\theta_{\rm opt}  &= \tan\angle \mathbf{f} \\
        & = -\frac{B_1(\cos2\phi-2)-B_2\sin2\phi}{B_2(\cos2\phi+2)+B_1\sin2\phi} \\
        & = \frac{\sin(2\phi+\psi)-2\sin\psi}{cos(2\phi-\psi)+2\cos\psi}.    
    \end{split}    
\end{equation}
As discussed in the previous section, the modulated phase delay will change the noise distribution, so the fingerprint direction will follow the changes together.
Figure.~\ref{fig:fingerprintangle2} displays how the angle of fingerprint varies with driving direction $\psi$ and modulated phase delay $\phi$. 

To simplify the model and for a convenient experimental setup, we carefully control the phase delay to zero $\phi = 0$, and the covariance matrix becomes
\begin{equation}\label{eq:covariance matrix with resonance_AP}
    \mathbb{C} = \frac{D\gamma^2 S_z^2}{8\Gamma_2} \begin{pmatrix}
        1 & 0 \\
        0 & 3
    \end{pmatrix}.
\end{equation}
In this case, the fingerprint direction is
\begin{equation}
    \label{eq:fingerprint in spin system_AP}
    \tan\theta_{\rm opt}  =\tan\angle \mathbf{f} = -\frac{B_1}{3B_2} = -\frac{1}{3\tan\psi},
\end{equation}
which is the result shown in the main text.

\section{THE GREEN'S FUNCTION IN THE SPIN SYSTEM}
\label{section_GreenFunction_ap}
The equation of transverse spin component $S_+$ under resonance conditions is given by Eq.(\ref{eq:Bloch_equations_in_rotating_frame}).
Re-expressing it in the form of SDE is as follows
\begin{equation}
    \frac{d \mathbf{S_+'}}{dt} = - \Gamma_2\mathbf{S_+'}+B_1(t)\mathbf{G_1}+B_2(t)\mathbf{G_2}+\mathbf{R}(t)
\end{equation}
where
\begin{align}
    \mathbf{G_1} &= 
    \begin{pmatrix}
        0\\\gamma S_z
    \end{pmatrix}, 
    \mathbf{G_2} =
    \begin{pmatrix}
        -\gamma S_z\\0
    \end{pmatrix}, \\
    \mathbf{R}(t) &= 
    \begin{pmatrix}
        -\gamma S_z \delta b(t) \sin(\omega t) \cos(\omega t)\\
        \gamma S_z \delta b(t) \cos^2(\omega t)  
    \end{pmatrix}.
\end{align}
The FPE of probability density is
\begin{widetext}
    \begin{equation}
        \begin{split}
            \frac{\partial P}{\partial t} &= - \nabla_S\cdot \left[ \left( -\Gamma_2 \mathbf{S_+'}+ B_1(t)\mathbf{G_1}+B_2(t)\mathbf{G_2} \right) P\right] + \frac{D \gamma^2 S_z^2 }{8}
            \begin{pmatrix}
                1 ,0 \\
                0 ,3 
            \end{pmatrix}
            :\nabla\nabla P \\
            &= \Gamma_2 \nabla_S\cdot(\mathbf{S_+}P) + \frac{1}{8} D \gamma^2 S_z^2(\partial^2_{S_x}+3\partial^2_{S_y})P-(B_1(t)\mathbf{G_1}+B_2(t)\mathbf{G_2})\cdot\nabla_S P
            \\
            &= \left(\mathcal{L}_0 + B_1(t)\mathcal{L}_1 + B_2 \mathcal{L}_2 \right) P,
        \end{split}
    \end{equation}        
\end{widetext}
where
\begin{align}
    \mathcal{L}_0^* &=\Gamma_2 \mathbf{S_+'}\cdot\nabla_S + \frac{1}{8}D\gamma^2 S_z^2(\partial^2_{S_x}+3\partial^2_{S_y})
    \\
    \mathcal{L}_1^* &=\mathbf{G_1}\cdot\nabla_S = \gamma S_z\partial_{S_y}
    \\
    \mathcal{L}_2^* &=\mathbf{G_2}\cdot\nabla_S = - \gamma S_z\partial_{S_x}
\end{align}
Then the stable probability density of $\mathcal{L}_0$ is
\begin{equation}
    P_0(\mathbf{S_+'}) =\frac{4\Gamma_2}{\sqrt{3}\pi D\gamma^2 S_z^2} \exp\left[-\frac{4\Gamma_2}{D\gamma^2 S_z^2} S_x'^2\right]\exp\left[-\frac{4\Gamma_2}{3D\gamma^2 S_z^2} S_y'^2\right]
\end{equation}
The Green's Function in the spin system can be expressed as:
\begin{align}
    \mathcal{G}_{S_+}^1(t) &=\Theta(t)\langle \mathcal{L}_{1}^* \mathbf{S_+}^{'(0)} \rangle =\Theta(t) {\gamma S_z}
    \begin{pmatrix}
        0 \\ 1
    \end{pmatrix}
    e^{-\Gamma_2 t},
    \\
    \mathcal{G}_{S_+}^2(t) &=\Theta(t)\langle \mathcal{L}_{2}^* \mathbf{S_+}^{'(0)} \rangle =\Theta(t) {\gamma S_z}
    \begin{pmatrix}
        -1 \\ 0
    \end{pmatrix}
    e^{-\Gamma_2 t},
\end{align}
and the observable $\mathbf{S_+'}$ can be written as
\begin{equation}
    \begin{split}
        \langle\delta S_+'(t) \rangle &= \int_{-\infty}^{t}d\tau \left[B_1(\tau) \mathcal{G}_{S_+}^1(t-\tau) + B_2(\tau) \mathcal{G}_{S_+}^2(t-\tau)\right]\\
        &= \frac{\gamma S_z}{2\Gamma_2}B_1(t) \hat{y} - \frac{\gamma S_z}{2\Gamma_2}B_2(t) \hat{x}         
    \end{split}
\end{equation}
which is consistent with the results of Eq.(\ref{eq:Signal_in_fingerprint}).Furthermore, according to the fluctuation dissipation theorem, the fluctuations of the transverse spin component(i.e. the projection noise in the spin system) can be derived from the Green's function
\begin{equation}
    \begin{split}
        \mathcal{S}_{\rm SPN} &= \coth(\frac{\beta\hbar \omega}{2})\Im\{i\mathcal{G}(\omega)\} 
        \\
        &= \frac{\gamma S_z}{\Gamma_2} \frac{1}{\omega^2 + \Gamma_2^2}    
    \end{split}
\end{equation}
Therefore, in experiments, the Green's function of the system can be determined by measuring the spin projection noise.

\section{EXPERIMENT SETUP DETAILS}
\label{section_experiment_ap}

The experiment setup is illustrated in the Fig.~\ref{fig:-EXPsetup}. 
To complete the fingerprint measurement, the phase of driving field and modulated Gaussian noise must be precisely controlled.
We use four synchronized lock-in amplifiers combined into a transverse magnetic field driver and signal detector.
The lock-in amplifier D mainly responsible for signal demodulation and acquisition.
Besides that, the lock-in amplifier D also generate the oscillation signal as an external reference to unify the driving frequency and phase on the other three lock-in amplifiers.  
As displayed in Fig.~\ref{fig:-EXPsetup}, the $\hat{x}$ and $\hat{y}$ coils are driven independently by two lock-in amplifiers A and B. 
By monitoring the voltage phase on the resistance at the end of the coil in situ, we control the phase of the transverse driving field through PID controller to generate a controllable linearly polarized magnetic field.
The lock-in amplifier C is set to generate the modulated Gaussian noise by demodulating a Gaussian white noise $\delta b(t)$ and output the in-phase component $\left\{\delta b(t)\cos\omega t\right\}$ of demodulation results.
However, the demodulation results have passed through a low-pass filter with a high bandwidth (indicated by $\left\{\cdots\right\}$, setting high bandwidth to reduce its weakening of noise $\delta b(t)$ assignments), 
which will inevitably introduce an additional delayed phase $\phi$ in the output, i.e.,
\begin{equation}
    \left\{b(t)\cos\omega t\right \} = \{b(t)\}\cos(\omega t+\phi),
\end{equation}
whose impact has been discussed in detail in the previous section.
The phase $\phi$ is mainly determined by the driving frequency, the filter bandwidth and the phase-frequency resonance of the coil.
Although the low-pass filter with large bandwidth(to pass the Gaussian noise) has almost no effect on noise amplitude, the impact of the additional delayed phase is significant.
To monitor and control this additional phase $\phi$, we use the modulation noise $\left\{b(t)\cos\omega t\right \}$ as input for demodulation with a narrow bandwidth of $200~\rm mHz$ on lock-in amplifier C,
and feedback the demodulation phase result to the driving channel to compensate the delayed phase $\phi$. 
In this way, we control the mean value of the phase delay $\phi$ becomes to zero and the fluctuating of approximately $100~\rm mdeg$, whose error is acceptable for experimental measurements.
By using PID feedback to control the output phase on three locking amplifiers, we can ultimately obtain the correct theoretical predicted measurement results discussed in the main text.

By measuring the Faraday rotation we can get the information of transverse spin polarization $S_x$ of Rb atoms. The voltage signal from balanced power detector (BPD) is
\begin{eqnarray}
V(t) = \eta S_x(t) = \eta (\tilde{S}_x(t)\cos \omega t-\tilde{S}_y(t)\sin \omega t),    \label{equ:six}
\end{eqnarray}
where $\eta$ is the spin to voltage coefficient determined by the gain and bandwidth of BPD, in our situation we consider it as a constant. 
The spin signal is detected by demodulated with the reference signal $V_r = \sqrt{2} e^{-i(\omega_ t+\theta)}$, therefore the in-phase and out-phase component $X$ and $Y$ can be expressed as
\begin{widetext}
    \begin{align}\label{equ:demodeResults}
        \begin{split}
            X &= \left\{\Re[V(t)\cdot V_{r}] \right\} = \left \{\frac{\sqrt{2}}{2}\eta \tilde{S}_x(\cos \theta +\cos (2\omega t - \theta)) - \frac{\sqrt{2}}{2}\eta \tilde{S}_y(\sin \theta +\sin (2\omega t - \theta))\right\},
            \\
            Y &= \left\{\Im[V(t)\cdot V_{r}]\right\} = \left \{\frac{\sqrt{2}}{2}\eta \tilde{S}_x(\sin \theta -\sin (2\omega t - \theta)) + \frac{\sqrt{2}}{2}\eta \tilde{S}_y(\cos \theta -\cos (2\omega t - \theta))\right\}.
        \end{split}
    \end{align}            
\end{widetext}
    
\begin{figure*}
    \includegraphics[scale=0.52]{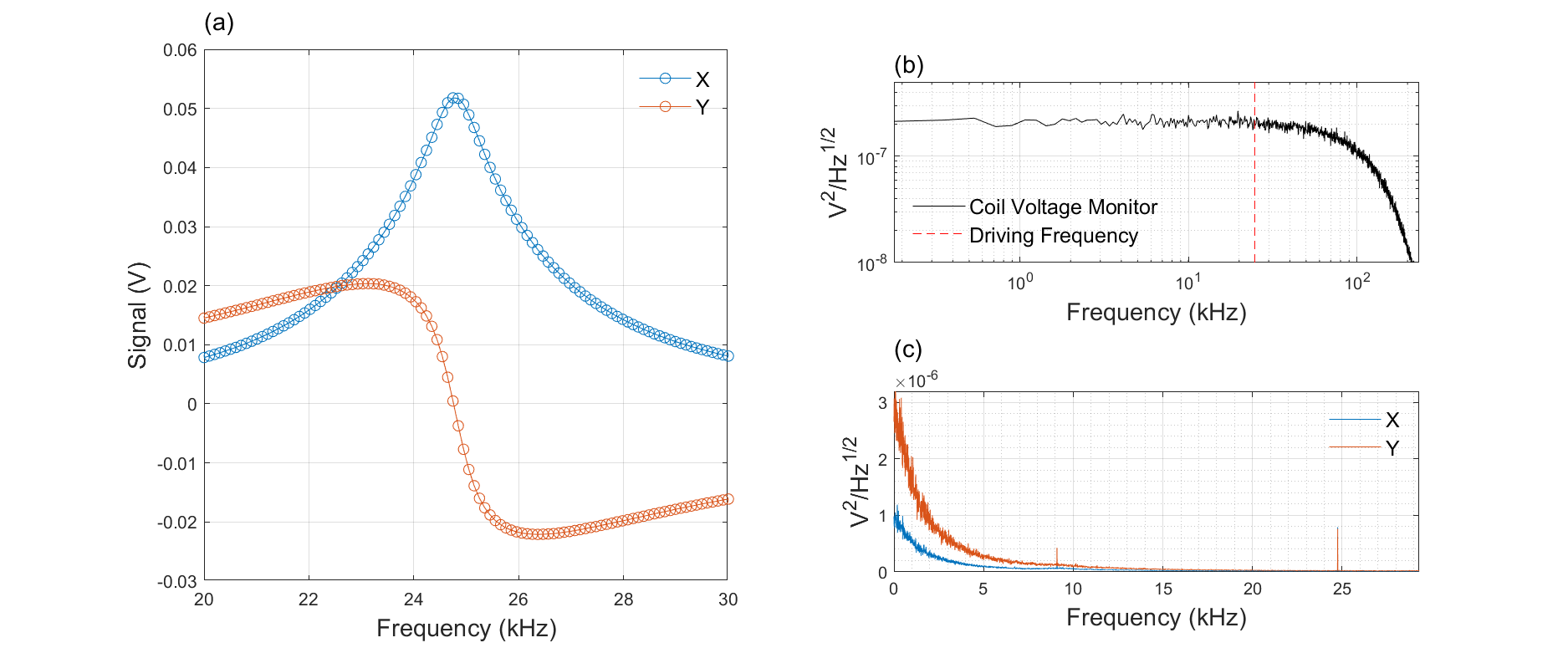}
    \caption{\label{fig:spectrum} RF spectrum and PSD of the noise. 
    (a) the RF spectrum of Rb atoms without noise.
    The drive direction is $\psi = \pi/2$ ($B_1 = 0~\rm nT,B_2 = 20~\rm nT$), and the demodulating phase $\theta = 0$, so the demodulation result $X$ and $Y$ is proportional to the spin signal $\tilde{S}_x$ and $\tilde{S}_y$
    In our system, the linewidth is $\Gamma_2 /(2\pi)\approx 1.5~{\rm kHz}$.
    (b) the PSD of modulated Gaussian noise $\delta b(t)\cos \omega t$ by monitor the voltages of the calibration resistance of x coil. 
    Owing to the frequency response of hardware devices, the noise power attenuates at high frequency. But is still a good white noise near the main frequency (red line on the figure)
    (c) the PSD of demodulation signal with $15~\rm kHz$ bandwidth of low-pass filter. The results meet Eq.~(\ref{equ:spinResPSD}).}
\end{figure*} 

The $\left\{\cdots\right\}$ indicates the operation of low-pass filtering.
As is discussed in Eq.~(\ref{equ:spinResPSD}), the spin signal $\tilde{S}_x, \tilde{S}_y$ mainly distributed at zero frequency with bandwidth within $\Gamma_2$, 
so by choose the low-pass filter bandwidth larger than $\Gamma_2$ and smaller than the driving frequency $\omega$, the multiplier frequency term in the output result can be easily eliminated.
In this case the demodulating result can be expressed as:

\begin{align}\label{equ:demodeResults}
    \begin{split}
        X &= \frac{\sqrt{2}}{2}\eta \left(\{\tilde{S}_x\}\cos \theta  +  \{\tilde{S}_y\}\sin \theta \right),
        \\
        Y &= -\frac{\sqrt{2}}{2}\eta \left(\{\tilde{S}_x\}\sin \theta  -  \{\tilde{S}_y\}\cos \theta\right),
    \end{split}
\end{align}    
where $\theta$ is the demodulating phase, which is also equal to the measurement direction in the rotating frame. 
The PSD of demodulation results are

\begin{eqnarray}\label{equ:demodePSD}
    \begin{aligned}
        G_{X}(\omega) =& \frac{1}{2}\eta^{2}\left(\{G_{\tilde{S}_{x}}\}\cos^{2}\theta +\{G_{\tilde{S}_{y}}\}\sin^{2}\theta  \right),
        \\
        G_{Y}(\omega) =& \frac{1}{2}\eta^{2}\left(\{G_{\tilde{S}_{x}}\}\sin^{2}\theta +\{G_{\tilde{S}_{y}}\}\cos^{2}\theta \right),
    \end{aligned}
\end{eqnarray}    
which also indicate the spin fluctuation frequency distribution on the detecting direction.

After all this theory analysis, we test our experiment system in the following aspects.
Firstly, we sweep the radio frequency (RF) spectrum of $^{87}\mathrm{Rb}$ to test the average part of spin signal, and the results display in Fig.~\ref{fig:spectrum}.
By setting the demodulating phase $\theta = 0$, the demodulation results is proportional to the spin signal $\tilde{S}_x$ and $\tilde{S}_y$.
According to Eq.~(\ref{eq:Signal_in_fingerprint}), the $\tilde{S}_x$ and $\tilde{S}_y$ are Lorentz and dispersion linearity, respectively, which the experiment results fit well.
The half height and half width of RF spectrum indicates $\Gamma_2/(2\pi )\approx 1.5~\rm kHz$.
After that, we apply the modulated Gaussian noise and try to estimate the noise frequency distribution in the spin signal. 
As shows in Fig.~\ref{fig:spectrum}(b), the correlation strength of the Gaussian noise can be calibrated by monitoring the voltage from calibration resistance of x coil. 
Although the noise power attenuates at high frequency owing to the frequency response of hardware devices, it can still be regarded as a white noise around the driving frequency.
To obtain the information of noise, we set the low-pass filter with $15~\rm kHz$ bandwidth, which is larger than $\Gamma_2/(2\pi)\approx 1.5~\rm kHz$ and smaller than the drive frequency $24.8~\rm kHz$. 
There is a Lorentz peak at around zero frequency with linewidth $\Gamma_2$, which is meet with Eq.~(\ref{equ:spinResPSD}). 
All key experimental parameters have been validated, and the results displayed in the main text is measured on this experiment system.

\bibliographystyle{apsrev4-2}

\end{document}